\documentclass[12pt]{article}
\usepackage[colorlinks,linkcolor=blue,citecolor=blue,bookmarks,bookmarksnumbered]{hyperref}
\usepackage{mathpazo,charter}
\usepackage[scaled=0.95]{helvet}
\usepackage{amsmath,XXXE}
\usepackage{graphicx,color}

 \SfTitles 
 \allowdisplaybreaks
\definecolor{gray}  {rgb}{0.60,0.80,0.80} 

\begin{document}

 \begin{center}
{\LARGE\sf\bfseries\boldmath
  On the Mirage of the Classical Electron\\[1mm] of Uhlenbeck and Goudsmit
 }\\*[5mm]
{\sf\bfseries A.\,P.~Batra
              and
              T.~H\"{u}bsch
}\\*[1mm]
{\small\it
      Department of Physics and Astronomy,\\[-1mm]
      Howard University, Washington, DC 20059
  \\[-4pt] {\tt\slshape  abatra@howard.edu {\rm and} thubsch@howard.edu}
 }\\[5mm]
{\sf\bfseries ABSTRACT}\\[3mm]
\parbox{150mm}{\addtolength{\baselineskip}{-2pt}\parindent=2pc\noindent
The idea that the electron is an extended charged object the spinning of which is responsible for its magnetic moment is shown to require a sizable portion of the electron to spin at speeds very close to the speed of light, and in fact to explode within an unacceptably short time, ${\sim}\,10^{-31}$\,s. The experimentally well-established magnetic moment of elementary particles such as the electron therefore must be accepted as an intrinsic property, with no need for classical models based on spatially extended objects. Emphasizing these facts in education, as early as possible, is important to the framing of the proper mind-set.
}
\end{center}
 \vglue5mm

\noindent
The physics approach to understanding Nature consists of a network of coherently constructed models that are proposed so as to reproduce concrete observations in Nature. Such models also tend to lead to new predictions, which subsequent experimentation is supposed to examine. When subsequent experimental results disagree with a model, it is evidently the latter that has to be modified if possible, or replaced. In this asymptotically improving process, it is important to realize that the predictions of any model must be followed out to their logical conclusion, and this is what we do here with the classical model of the electron.

The Abraham-Lorentz model of the electron|pictured variably as a spherical shell or as a homogeneous solid ball with the electron's charge distributed uniformly|was introduced in 1902 by Max Abraham\cite{rMA-Electron} and revisited in 1904 by Hendrik Lorentz\cite{rHL-Electron}. If such a spatially extended object was spinning, the spinning charge would create a circular current, which in turn would create a magnetic dipole field. The intrinsic magnetic dipole moment of the electron was established and measured by George Eugene Uhlenbeck and Samuel Abraham Goudsmit in 1925\cite{rUG-Spin}. It was thus attributed to the spinning of such a charged and spatially extended object, providing the value of $\inv2\hbar$ for the magnitude of the spin angular momentum of the electron.

Note that the so-attributed spinning of the electron is a reverse-engineered quantity, hinging on the Abraham-Lorentz model of the classical electron. The only unambiguously established physical property of the electron in this respect is its intrinsic magnetic dipole moment.

Let's assume that the electron has some spatial extension but otherwise an unknown shape, and let $r_e$ be the minimal radius of a sphere that completely encloses such an electron.
The energy of the electron's electric field (using only the monopole term in the multipole expansion of the $\vec{E}$-field) is then 
\begin{equation}
  E_e = \inv2\int\rd^2\vec{r}~\vec{E}^{\,2} = \frac12\,\frac{e^2}{4\p\e_0}\,\frac1{r}
      = \frac{\a_e\hbar c}{2\,r}
  \qquad\text{for }r \geq r_e~,
 \label{e:Energy}
\end{equation}
where $\a_e=\frac{e^2}{4\p\e_0\,\hbar c}\approx\frac1{137.036}$ is the electromagnetic fine structure constant.
Assuming that the mass\ft{We adhere to the nomenclature, championed by Prof.~Okun\cite{rLO-Mass}, that {\em\/mass\/} is a relativistic invariant and its $c^2$-multiple equals the {\em\/rest energy\/} of an object; this then equals the expression\eq{e:Energy}.} of the electron stems entirely from this energy, we identify $E_e=m_e c^2$ and have
\begin{equation}
  r_e = \frac{\a_e}2\frac{\hbar}{m_ec}\qquad\text{or}\qquad
  r_e = \a_e\b\frac{\hbar}{m_ec} = \b\cdot2.81794\times10^{-15}\,\text{m}
 \label{e:radius}
\end{equation}
for the radius of the electron. Note that the oft-quoted result neglects the numerical factor of $\b=\inv2$, on grounds of offering estimates rather than precise results: after all, we have neglected all the multiple contributions to $\vec{E}$ beyond the spherically symmetric monopole term.
 In addition, we have used here the relativistic relationship between the rest-energy, the mass and the speed of light; before 1905, several similar formulas were being proposed by various researchers of the time, differing only in the numerical proportionality factors that were however all of order $\,{\sim}1$.
 Herein, we lump these variations of the model in the numerical parameter $\b\,{\sim}\,1$.

In turn, given that the electron is assumed to have some spatial extension, its moment of inertia for spinning about an axis that passes through its center-of-mass must be
\begin{equation}
  I_e = \x\, m_e r_e^{~2},
 \label{e:MomIn}
\end{equation}
where $\x$ is a numerical constant not greater than 1, and equals to 1 if all the mass of the electron is concentrated at the distance $r_e$ from the axis of rotation. Being that we expect some of the mass to be distributed also at smaller distances, it follows that $\x\leq1$. For a spherical shell, $\x=\frc23$, whereas for a solid spherical ball, $\x=\frac25$. 

Anticipating that the rotation might be relativistic, the angular momentum of the spinning electron then has a magnitude of
\begin{equation}
  S_e = |\vec{r}\times\vec{p}| = \x(r_e)(m_e\g v_t)
      = I_e\g\w = I_e\g\frac{v_t}{r_e},
 \label{e:AngMom}
\end{equation}
where $v_t$ is the tangential speed of rotation of the electron, at the distance of $r_e$ from the center-of-mass, and $\g=(1-(v_t/c)^2)^{-1/2}$. (By 1925, the special theory of relativity was fairly well accepted, and certainly amongst theorists.)
 Combining equations\eq{e:AngMom} and\eq{e:radius}, we have
\begin{equation}
  S_e = \a_e\b\x\,\hbar\,\g\frac{v_t}{c}
\end{equation}
which ought to be identified with $S_e=\inv2\hbar$, the value cited by Uhlenbeck and Goudsmit. Doing so however immediately implies that
\begin{equation}
  S_e = \a_e\b\x\,\hbar\,\g\frac{v_t}{c} = \inv2\hbar,
   \qquad\ie,\qquad
  \frac{v_t}{c} = \frac{\sqrt{1-(v_t/c)^2}}{2\a_e\b\x},
 \label{e:v/c0}
\end{equation}
which is easily solved to produce\ft{In fact, H.\,A.~Lorentz was suspicious of the idea of spin, having estimates $v_t\sim10c$\cite{rU-50yS}, which must mean that he used|ironically|a non-relativistic expression for angular momentum.}
\begin{equation}
  \frac{v_t}{c} = \frac1{\sqrt{1+(2\a_e\b\x)^2}}
  ~\gtrsim~ 0.999894,
 \label{e:v/c}
\end{equation}
and this {\em\/lowest\/} value is obtained by using the maximal values of $\b,\x\lesssim1$. As noted, for most shapes the monopole term in the expansion of $\vec{E}$, and so also of the expression\eq{e:Energy}, are not likely to be overwhelmed by the higher order contributions in the multipole expansion.

So, Uhlenbeck and Goudsmit's reverse-engineered attribution of the {\em\/measured\/} intrinsic magnetic dipole moment to a {\em\/fictitious\/} spinning requires the tangential speed, $v_t$, of the electron to be larger than 99.9894\% of the speed of light in vacuum! 

But, there's more: Being that the roughly spherical distribution of the electron's charge is rotating, it moves in an accelerated fashion, and therefore radiates. This radiation loss of course depletes the energy content|and so the rest mass|of the spinning electron, and may be estimated using the relativistic version of the Larmor formula obtained by Alfred-Marie Li\'enard in 1898\cite{rLL2,rJDJ}:
\begin{align}
  \frac{\triangle E}{\triangle t}
  &= \frac23\frac{e^2}{4\p\e_0}\frac1{c}\g^6
      \Big[\big(\frc{a_r}c\big)^2-\big(\frc{v_t}c\,\frc{a_r}c\big)^2\Big]
   = \frac23\frac{e^2}{4\p\e_0}\frac{a_r}{c^3}\g^6
      \Big[1-\big(\frc{v_t}c\big)^2\Big]
   = \frac23\frac{e^2}{4\p\e_0}\frac{a_r^{\,2}}{c^3}\g^4,\nn\\
  &= \frac23\frac{\a_e\,\hbar}{c^2}a_r^{\,2}\g^4
   = \frac23\frac{\a_e\,\hbar}{c^2}\frac{\big(v_t^{\,2}/r_e\big)^2}{\big(1-(v_t/c)^2\big)^2}
   = \frac23\a_e\,\hbar c^2\,\Big(\frac1{r_e}\Big)^2\,
      \frac{\big(v_t/c\big)^4}{\big(1-(v_t/c)^2\big)^2}.
\end{align}
Using the above results, \Eq{e:v/c} for the ratio $v_t/c$ and\eq{e:radius} for the radius of the electron, $r_e$, this becomes
\begin{equation}
   \frac{\triangle E}{\triangle t}
   ~=~ \frac23\a_e\,\hbar c^2\,\Big(\frac{m_ec}{\a_e\b\hbar}\Big)^2
        \frac1{(2\a_e\b\x)^4}
   ~=~ \frac1{24}\,\frac{m_e^{\,2}c^4}{\a_e^{\,5}\,\b^6\,\x^4\,\hbar}.
\end{equation}
This estimate may be used to determine how fast it would take for the electron to radiate away all of its rest mass $\triangle E\approx m_ec^2$:
\begin{equation}
  \triangle t
  =\frac{(\triangle E=m_ec^2)}
        {\frac1{24}\,\frac{m_e^{\,2}c^4}{\a_e^{\,5}\,\b^6\,\x^4\,\hbar}}
  =24\,\frac{\a_e^{\,5}\,\b^6\,\x^4\,\hbar}{m_e\,c^2}
  ~\approx~6.38506\,{\times}\,10^{-31}\,\text{s},
 \label{e:Dt}
\end{equation}
where we have again used that $\b,\x\lesssim1$.
 This is clearly an unacceptable result: during this amount of time, the electromagnetic radiation|through which the electron is supposed to lose all of its mass|can traverse at most $1.91552\times10^{-22}$\,m, which is only a minuscule fraction ($\sim\fRc{\SSS1}{\SSS14{,}704{,}800}$) of the electron's radius\eq{e:radius}.

 Indeed, the actual time would have to be {\em\/longer\/} than this, as the estimate\eq{e:Dt} does not take into account that as the mass of the electron reduces through radiation loss, the radius of the electron grows; see\eq{e:radius}. For the angular momentum\eq{e:AngMom} to remain conserved, the tangential speed then must also reduce. In turn, the radial acceleration $a_r=v_t^{\,2}/r_e$ then reduces both because the tangential velocity reduces and because the radius of the electron grows, whereupon the radiation loss reduces even more, depending on the square of the reduced radial acceleration. Although this significantly increases this estimate of $\triangle t$, it does not seem plausible that this could extend the effective lifetime of the electron\eq{e:Dt} by the 64 orders of magnitude or more needed to agree with the experimental bound\cite{rPDG}.
 The model also predicts that the radius of the electron grows unboundedly as it loses its mass---contrary to all observations.

 Of course, by 1925, it was known that the orbiting electrons in a Hydrogen atom do not radiate, and a similar {\em\/ad hoc\/} Bohr-like postulate could be invoked to prevent the spinning electron from exploding. However the situation here is significantly different from Bohr's model of the Hydrogen atom, where the Coulomb force balances the electron in its stable orbit: there is {\em\/nothing\/} in the Abraham-Lorentz model of the spinning electron that holds it contained! 

The above critical analysis of the Abraham-Lorentz-Uhlenbeck-Goudsmit model of the spinning electron quite evidently has rather preposterous implications. Besides violating the fact of Nature that electrons are stable particles (no electron has ever been observed to decay\cite{rPDG}), this exploding electron model also de-localizes the electric charge of the electron.

Nevertheless, physicists at the time accepted Uhlenbeck and Goudsmit's reverse-engineered attribution of the {\em\/measured\/} intrinsic magnetic dipole moment to a {\em\/fictitious\/} spinning of the electron. Even the subsequent realization that the magnetic moment of elementary particles such as the electron is an intrinsic property|just as is its mass and electric charge|the name ``spin'' unfortunately stuck! Generation after generation of students thus become misguided by the ``spin'' as an ``explanation'' of the magnetic moment of the electron.

Whereas history cannot be undone and the name ``spin'' is too entrenched in the physics jargon to be eradicated, we hope that at least the educators will agree to include this cautionary note in their contributions to the forming of the mindset of the coming generations. In our teaching experience, this is effectively accomplished by emphasizing the reverse-engineered nature of the notion of ``spin,'' derived from the observed magnetic dipole moment.

\paragraph{\bfseries Acknowledgments:}
We should like to thank Profs.\ W.\,P.~Lowe, J.\,H.~Stith and D.\,D.~Venable for critical reading of the manuscript and constructive suggestions.
TH is grateful to the Department of Energy for the generous support through the grant DE-FG02-94ER-40854, as well as the
 Physics Departments at the University of Central Florida, Orlando FL,
and at the
 Faculty of Natural Sciences of the University of Novi Sad, Serbia,
 for recurring hospitality and resources.

\raggedright
\end{document}